\documentclass[%
aps,
preprint,
amsmath,amssymb,
]{revtex4-2}

\usepackage{graphicx}
\usepackage{dcolumn}
\usepackage{subfigure}
\usepackage{fancyhdr}
\usepackage{lastpage}
\usepackage{graphicx, wrapfig, setspace, booktabs}
\usepackage[english]{babel}
\usepackage{hyperref}
\usepackage{url, lipsum}
\usepackage{epstopdf, epsfig}
\usepackage{amsmath,amsfonts,amssymb}
\usepackage{float}
\usepackage{gensymb}
\usepackage{mathrsfs}
\usepackage{url}
\usepackage{natbib}

\begin{document}


\title{Unpredictable tunneling in a retarded bistable potential}

\author{Álvaro G. López$^{1}$}\email{alvaro.lopez@urjc.es}
\author{Rahil N. Valani$^{2}$}

\affiliation{$^1$ Nonlinear Dynamics, Chaos and Complex Systems Group.\\Departamento de F\'isica, Universidad Rey Juan Carlos, Tulip\'an s/n, 28933 M\'ostoles, Madrid, Spain} 
\affiliation{$^2$School of Mathematical Sciences, University of Adelaide, Adelaide, South Australia 5005, Australia
3}


\date{\today}

\begin{abstract}
We have studied the rich dynamics of a damped particle inside an external double-well potential under the influence of state-dependent time-delayed feedback. In certain regions of the parameter space, we observe multistability with the existence of two different attractors (limit cycle or strange attractor) with well separated mean Lyapunov energies forming a two-level system. Bifurcation analysis reveals that, as the effects of the time-delay feedback are enhanced, chaotic transitions emerge between the two wells of the double-well potential for the attractor corresponding to the fundamental energy level. By computing the residence time distributions and the scaling laws near the onset of chaotic transitions, we rationalize this apparent \emph{tunneling-like effect} in terms of the \emph{crisis-induced intermittency} phenomenon. Further, we investigate the first passage times in this regime and observe the appearance of a Cantor-like fractal set in the initial history space, a characteristic feature of hyperbolic chaotic scattering. The non-integer value of the uncertainty dimension indicates that the residence time inside each well is unpredictable. Finally, we demonstrate the robustness of this tunneling intermittency as a function of the memory parameter by calculating the largest Lyapunov exponent.
\end{abstract}

\maketitle
\newpage


\section{Introduction}
Empirical discoveries in hydrodynamic quantum analogs with walking droplets during the last two decades have demonstrated that properties once believed to be exclusive to the microscopic realm of physics can also be observed in macroscopic systems~\cite{cou05,pro06}. Some of these include orbit quantization~\cite{for10}, diffraction and interference phenomena through slits~\cite{cou06,and15,puc18}, tunneling over barriers~\cite{edd09,bush20}, and non-classical correlations of particles~\cite{pap22,val18b}. Walking droplet experiments consist of a silicone oil droplet bouncing and walking in resonance with an underlying vibrating fluid bath~\cite{cou05}. The slowly decaying waves produced by the repeated bounces endow the particle with a time-delayed self-force, introducing memory effects in the system \cite{oza13,bus18}. Specifically, using submerged barriers in the fluid, the importance of chaotic dynamics to understand the tunneling effect in pilot-wave systems has been highlighted in recent works~\cite{bush20}.

Memory produced by time-delays is frequently discarded when studying problems in the atomic realm of physics, where complex electrodynamic interactions between particles navigating in a fluctuating vacuum~\cite{del14} are simplified to a static Coulombian potential~\cite{bohr13,sch26}. Theoretical works on classical electrodynamics of extended bodies~\cite{lop20,lop21} have shown the existence of self-sustained oscillations with \emph{zitterbewegung} frequency, triggered by state-dependent time-delayed self-interactions. But even in systems of interacting charged point masses, the retarded Li\'{e}nard-Wiechert potentials~\cite{lie98} can produce unexpected nonlinear phenomena~\cite{raj04,sil23}. The expression Raju-Atiyah's hypothesis has been coined to denote this necessity of taking into account time retardation in the differential equations of motion of elementary charged particles. Briefly put, this paradigm proposes that functional differential equations are behind some of the effects that we observe at the atomic scale~\cite{lop23}. 

The importance of time-delay has been extensively emphasized in a broad range of scientific disciplines. Apart from fundamental physics, we can find them in chains of chemical reactions~\cite{sc86}, cardiac oscillations~\cite{mac77}, propagation of impulses through the nervous system~\cite{han22} or modeling of the cell cycle~\cite{fer11}. Time delays are also important in understanding climate behavior, for example the El Niño-Southern Oscillation~\cite{bou07}. Moreover, in epidemiology and population dynamics, memory effects produce unexpected phenomena~\cite{sal98}. Even in the modeling of economic cycles time-delay might play an essential role~\cite{kal35}. Nevertheless, dynamical systems with varying time-delay have been studied less in comparison, primarily due to their inherent mathematical and dynamical complexity~\cite{ins07,mar15,mul18,raj15}. 

In the present work, we propose a novel nonlinear retarded oscillator to investigate the chaotic dynamics of tunneling-like phenomenon in an external double well potential; similar to the well initially used by F. Hund to understand molecular spectra, where the electrons can transit between the several atomic regions that comprise the molecule~\cite{hun27}. The paper is organized as follows: We start by presenting the dynamical system in Sec.~\ref{sec: DE}. We then in Sec.~\ref{sec: Quantization} use standard numerical tools from nonlinear dynamics to uncover coexisting attractors, such as quantized orbits~\cite{lop23} and chaotic attractors, at the same parameter values with different mean Lyapunov energies. By systematically varying the memory of the system, we explore several regimes of chaotic intermittency at the fundamental energy level. In Sec.~\ref{sec: Intermittency}, the tunneling-like behavior in the intermittency regime is characterized through computation of the residence time distribution inside one well. We also analyze the scaling law, which typically describes the critical slowdown close to the onset of intermittent behavior. In Sec.~\ref{sec: Tunneling}, the inherent unpredictability of such intermittent transitions is shown by computing the fractal dimension of the Cantor-like set underlying the chaotic dynamics. In Sec.~\ref{sec: Robustness} we test the robustness of this tunneling-like intermittent dynamics between the two wells by computing the largest Lyapunov exponent in the parameter space. We discuss and conclude in Sec.~\ref{sec: Discussion}.

\section{Dynamical equations}\label{sec: DE}

We consider a particle with mass $m$ subjected to a linear drag force with damping coefficient $\zeta$. This particle is subjected to an external quartic bistable potential $V(x)=-\alpha x^2/2 + \beta x^4/4$, along with a self-interaction quadratic retarded potential $Q(x_{\tau})=\lambda x_{\tau}^2/2$ with state-dependent time-delay, where $x_{\tau}=x(t-\tau(x))$. Following studies in classical electrodynamics, the concept of a retarded potential~\cite{lie98,lop23} is borrowed hereafter to denote this state-dependent time-delay. They are akin to the Li\'enard-Wiechert potentials, which have been related to the Bohmian quantum potential in studies of electrodynamic bodies~\cite{lop20}. For small state-dependent time-delays, these nonlinear oscillators can be approximated as Li\'enard systems~\cite{lop23}. Therefore, we can write the equation of motion for the particle, which takes the form of a nonlinear time-delayed oscillator, as 
\begin{equation}
    m\ddot{x}+\zeta \dot{x} + \dfrac{dV}{dx}+\dfrac{dQ}{dx_{\tau}} = 0. 
    \label{eq:1}
\end{equation}
For simplicity, and without loss of generality, we fix the parameters $m=1$, $\zeta=0.1$, $\alpha=1$ and $\beta=0.1$, unless otherwise stated. This model is an extension of the self-excited oscillator developed in Ref.~\cite{lop23} that considered an external harmonic potential, which in turn was inspired by another nonlinear oscillator encountered in the study of the dynamics of extended electromagnetic bodies~\cite{lop20}. These particles experience self-forces, which result in self-oscillations through the well-known Hopf bifurcation~\cite{lop21,jen13}. To some extent, we can qualitatively interpret the retarded potential as the result of complex self-interactions of the particle with its history as it evolves inside the double-well potential. In this manner, even though the particle can be localized in one well at a given time, it can interact with its past motion on the other side of the well, entailing the use of the tunneling concept. 
\begin{figure}
\centering
\includegraphics[width=0.5\columnwidth]{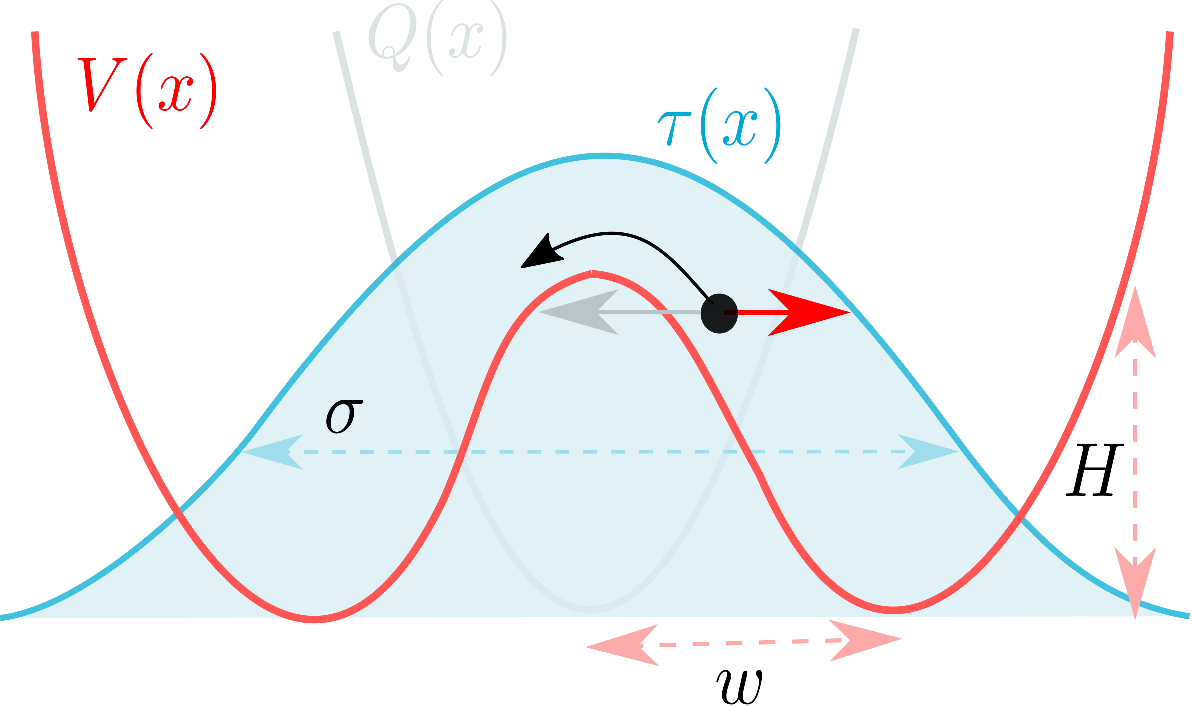}
\caption{\textbf{Mathematical model}. A particle of mass $m$ (black dot) moving inside an external double-well $V(x)$ subjected to a harmonic retarded potential $Q(x_{\tau})$ with $x_{\tau}=x(t-\tau(x))$. The state-dependent time-delay $\tau(x)$ having a Gaussian profile with standard deviation $\sigma$ is shown in blue. This Gaussian delay profile is centered at the local minimum of the quadratic well $Q$. This time-delay non-conservative force allows the particle to jump over the central barrier of the double well. The width $w$ and height $H$ of the potential are depicted as well.}
\label{fig:1}
\end{figure}

We note some points regarding the choice of the delay function $\tau(x)$ for this study. Some restrictions on this function must be provided to ensure that the system is mathematically and physically well-behaved. Firstly, we want the trajectories to remain bounded in the external well as $x \rightarrow \pm \infty$ , so that the state-dependent delay decays to zero asymptotically. It is also reasonable to demand that the delay function remains bounded over its entire domain, ensuring that the feedback coming from the past does not diverge. In this way, we limit the memory of our oscillator to a finite domain of its past. Thirdly, we preserve the symmetry with respect to spatial reflections about the center of the external double-well potential, as also done in our previous work with an external harmonic potential~\cite{lop23}. Hence, we preserve the reflection symmetry in the model noting that this symmetry may be spontaneously broken in the dynamics. A simple transcendental function that accounts for these three requirements stated above is the Gaussian distribution~\cite{raj15,lop23}. Hence, we choose the delay function to be $\tau(x)=\tau_0 e^{-x^2/2 \sigma^2}$, and fix $\sigma=7/\sqrt{2}$ hereafter. The parameter $\tau_0$ represents the maximum value of the time-delay feedback, attained at the local maximum of the double-well potential. Together with $\lambda$, it constitutes a key parameter investigated in the present study to control the effects of memory. A similar state-dependence of the time-delay has been previously used in the literature to investigate the effects of memory in vibrational resonance~\cite{raj15}.
\begin{figure}
\centering
\includegraphics[width=0.95\columnwidth]{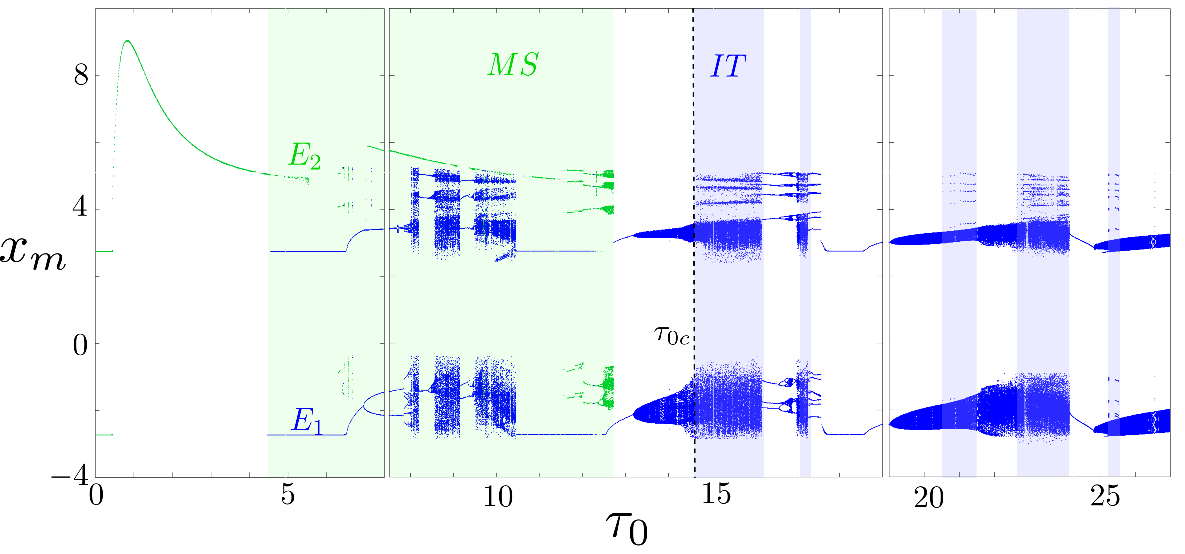}
\caption{ \textbf{Bifurcation diagram.} A bifurcation diagram of the maxima map $x_m$ is shown as a function of the memory parameter $\tau_0$. Different initial histories have been chosen and the corresponding asymptotic sets are represented using two different colors (blue and green), whenever they exist. Multistable (MS) region (green background) can be identified where two different asymptotic states exist. The regions where intermittent (IT) behavior has been observed are shown with a blue background, and the critical value $\tau_{0c}$ has been marked (dashed line). Several routes to chaos, including period doubling and quasiperiodic cases~\cite{spr08}, are evident from the bifurcation diagram.}
\label{fig:1}
\end{figure}

We consider the phase space representation of our dynamical system by using the phase-plane formed by the dynamical variables $x$ and $\dot{x}$, as it is conventionally done in the study of nonlinear dynamical systems, especially with mechanical and electronic oscillators. However, we remark that the true phase-space of our dynamical system is infinite-dimensional due to the presence of time-delay. Thus, initial history functions have to be provided to integrate the Eq.~\eqref{eq:1}, instead of mere initial points in the $(x,\dot{x})$ phase plane. In this phase-plane representation, we can write the differential equation as follows
\begin{align}
\dot{x} & = y, \label{eq:2} \\
\dot{y} & = - 0.1 y + x - 0.1 x^3 - \lambda x_{\tau}, 
\label{eq:3}
\end{align}
which can be integrated numerically using a residual control integrator \cite{sha05}.
\begin{figure}
\centering
\includegraphics[width=0.5\columnwidth]{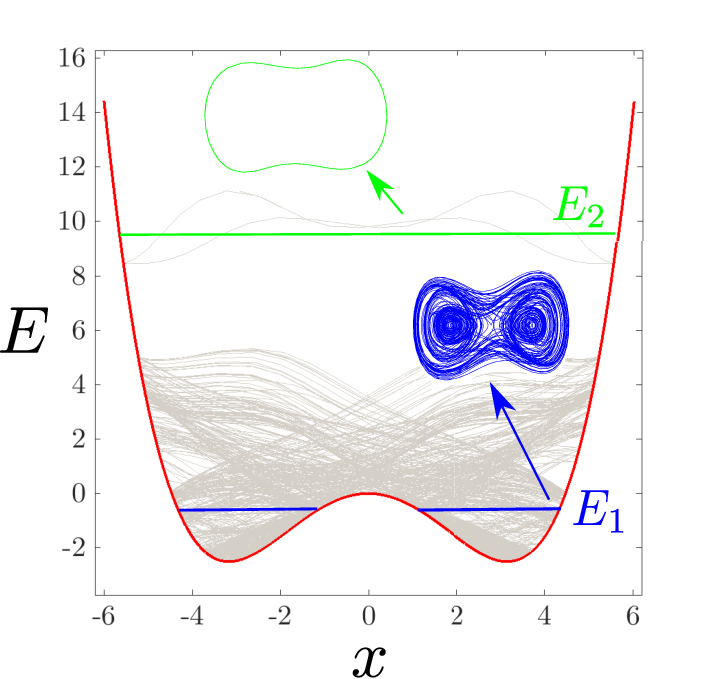}
\caption{\textbf{Energy levels}. A two-level system of distinct attractors with quantized energies for $\tau_0=8.1$. At the fundamental level is a chaotic attractor with average value of Lyapunov energy function $E_1$ (blue), while the higher energy level is a periodic limit cycle with average value of Lyapunov energy function $E_2$ (green). The double-well potential is represented in red. The time evolution of the Lyapunov energy function is shown in grey, which depicts considerable fluctuations, but the two energy levels are well resolved.}
\label{fig:2}
\end{figure}

\section{Orbit quantization}\label{sec: Quantization}

We now investigate the routes to chaotic dynamics for the dynamical system represented by Eqs.~\eqref{eq:2} and \eqref{eq:3}. For this purpose, bifurcation diagrams have been numerically computed as a function of the memory parameter $\tau_0$. We set a value of $\lambda=0.25$, which is consistent with previous works~\cite{lop23}, and for which the existence of quantized orbits and the phenomenon of intermittency are more clearly appreciated. A typical bifurcation diagram of the system is shown in Fig.~\ref{fig:1}. To represent the bifurcation diagram, the system of Eqs.~\eqref{eq:2} and \eqref{eq:3} has been solved for three different initial histories using an inbuilt residual control integrator in MATLAB. Trajectories in the range $t \in [0, 3000]$ are computed for each memory parameter value $\tau_0$, and long transients, up to $t = 2400$, have been discarded, since delayed dynamical systems frequently exhibit supertransients~\cite{lak11}. The maxima map of the spatial coordinate $x$, after discarding transients, is plotted and colored according to the asymptotic attractor approached for each value of the memory parameter.  The maxima (minima) map is computed by selecting the local maxima (minima) of the time series of a dynamical system.
\begin{figure}
\centering
\includegraphics[width=1.0\columnwidth]{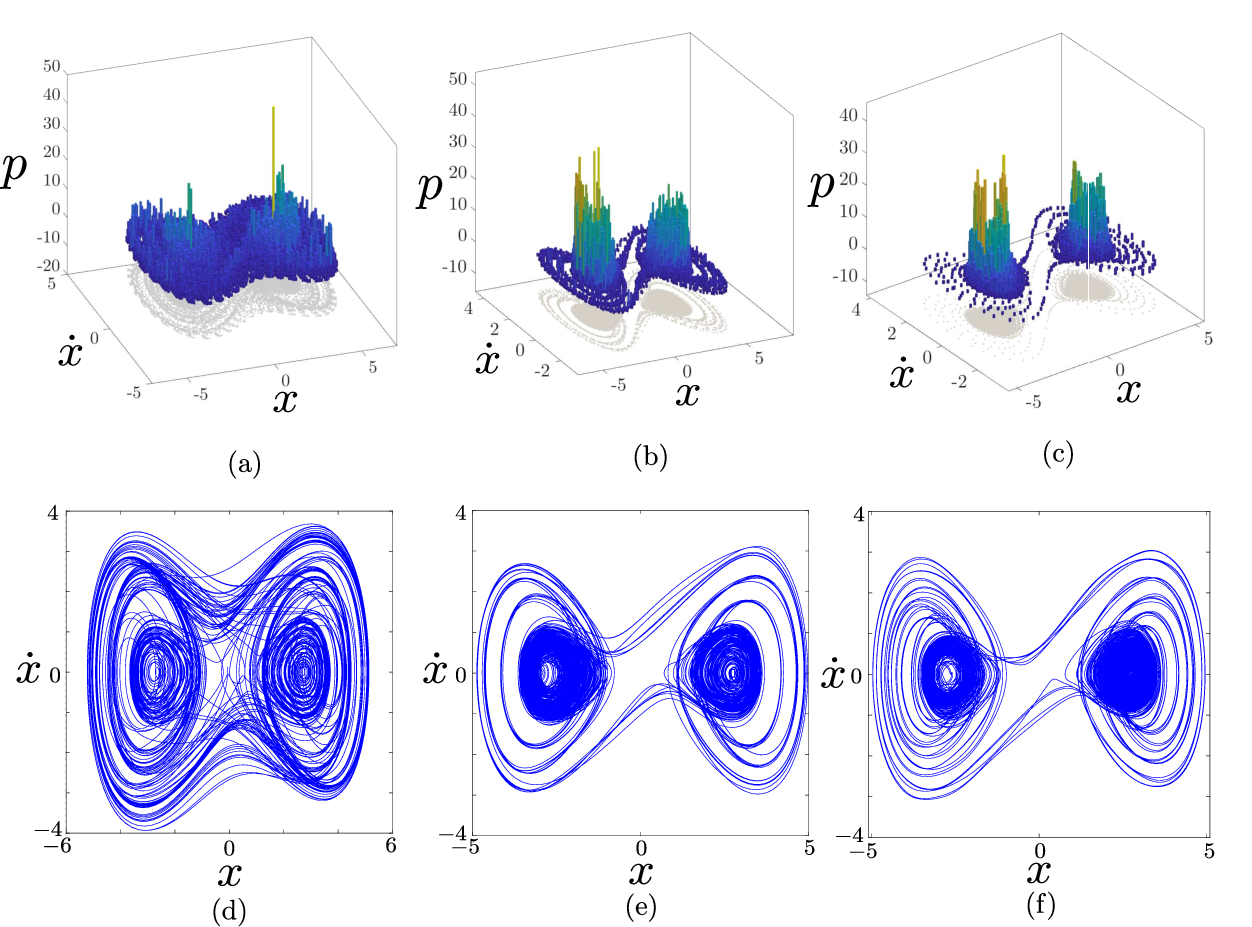}
\caption{\textbf{Invariant measures}. The phase-space frequency $N$ of the trajectories in the intermittent regimes at different value of memory parameter (top panel), together with their corresponding phase-plane trajectories (bottom panel). (a,d) For the memory parameter value $\tau_0=8.1$, where the particle spends large parts of the trajectory exploring both wells. (b,e) For the memory parameter value $\tau_0=15.0$, where the particle spends most of the time deep inside one well, and rapidly transitions to the other well. (c,f) For the memory parameter $\tau_0=24.0$, where the time spent within the wells is the largest and the transition between wells the shortest.}
\label{fig:3}
\end{figure}

Following previous studies, we have considered the most natural and useful choice of initial history functions as periodic solutions~\cite{lop23}. Physically, this corresponds to periodically forcing the particle up to the present time and then letting it evolve according to its equation of motion. Therefore, we take the history function of the form $x(t) = A \sin(\omega t + \varphi)$ for $t<t_0$, with $t_0=0$. To detect multistable parameter regimes when they exist, for example represented by two or more coexisting stable limit cycles, we use three different initial history functions having different values of $A$, $\omega$ and $\varphi$. Two initial conditions are set inside the respective wells for $A= \pm 1$, $\omega=0.5$, and $\varphi=0.1$. The third one is used to capture large amplitude limit cycles for $A= 13.5$, $\omega=0.5$ and $\varphi= 6\pi/5$. Other choices of initial conditions produce similar results. Systematic numerical explorations reveal that these three initial histories suffice to represent the two attractors that we have found at these parameter values of the dynamical system. 
\begin{figure}
\includegraphics[width=1.02\columnwidth]{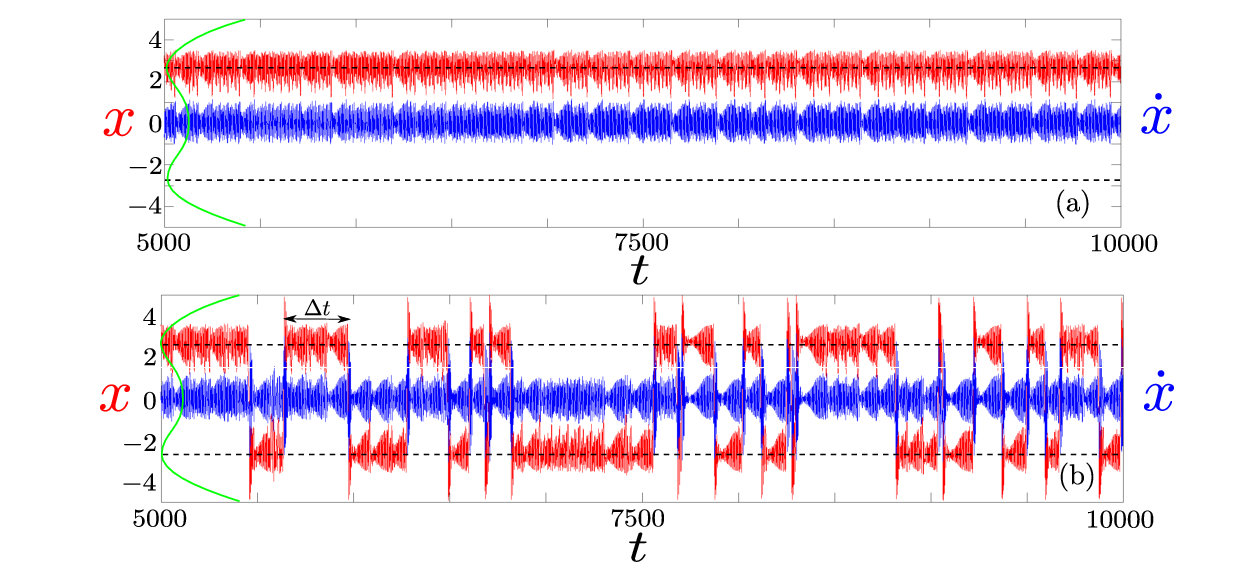}
\caption{\textbf{Transition to intermittency}. The time series for a particle initiated in the left well, before and after the crisis. (a) We see a confined particle, oscillating chaotically in one of the two wells, for $\tau_0=14.5$. (b) For $\tau_0=15.0$, the intermittent behaviour is evident, with the particle spending intervals of unpredictable residence time $\Delta t$ in each well, and switching rapidly to the other well.}
\label{fig:46}
\end{figure}

In the absence of self-interactions ($\lambda = 0$), the damped particle asymptotically approaches one of the two symmetric minima of the double well, where the Lyapunov energy function, $E(x,\dot{x})=\dot{x}^2/2+V(x)$, takes its minimum value. However, including self-interaction by increasing the value of $\lambda>0$, the effects of the retarded potential come into play. Then, as we increase the maximum memory of the system $\tau_0$, we observe that the fixed points migrate from the minima of the double well potential towards the maxima of the potential at $x=0$. This is followed by a Hopf bifurcation occurring at $\tau_{0}=0.42$ that produces an extended limit cycle encompassing both wells (green in Fig.~\ref{fig:1}). Further increase of the memory leads to the birth of two new symmetric fixed points inside their respective wells (blue in Fig.~\ref{fig:1}), which coexist with the previous limit cycle. The multistable (MS) region begins at the birth of these two symmetric fixed point and the system now has two separated energy levels, distinguished via the average value of the Lyapunov energy function $\bar{E}$, corresponding to the two attractors in this MS region. As the memory parameter is increased further in the MS region, we see a route to chaotic dynamics for the newly born fixed points (blue). This chaotic attractor constitutes the first energy state where the trajectory switches erratically between the minima of the two wells. Conversely, the second energy state corresponding to the extended limit cycle (green) remains periodic with increases in the memory parameter, however the size of the limit cycle decreases. 

In Fig.~\ref{fig:2} we have shown the time average value of the Lyapunov energy function for both attractor states, which are well-resolved in the energy diagram. The energy has been computed using the Lyapunov function and the average has been integrated using the well-ḱnown trapezoidal rule. For the lower energy level $E_1$, the average energy value is below the local maximum of the double well potential. However, we observe considerable oscillations above this energy level, showing that the system is non-conservative. As it has been highlighted in previous works, these self-excited open systems can be regarded as thermodynamic engines oscillating far from equilibrium~\cite{jen13,lop232}. In this respect, retarded nonlinear oscillators are very different from conservative Hamiltonian systems; for the latter, the energy functional is a constant of motion and such systems are endowed with a canonical symplectic topology. 

As the memory parameter is increased beyond $\tau_0=10$ in the MS region in Fig.~\ref{fig:1}, we encounter two chaotic crises phenomena. The first crisis destroys the chaotic attractor at the fundamental energy level (blue) at $\tau_0=10.44$, while the second crises destroys the chaotic attractor at the higher energy level (green). Once the chaotic attractor at the higher energy level is destroyed near $\tau_0=12.56$, the MS region ceases. Beyond this regime for larger values of the memory parameter, two new limit cycles, one in each well, are born (blue) near $\tau_0=12.56$, which further destabilize through a second Hopf bifurcation at $\tau_0 = 13.22$, rendering two quasiperiodic attractors, one inside each well. As suggested by time series analysis and the computation of the largest Lyapunov exponent, these two symmetric quasiperiodic attractors later become chaotic. Finally, for an approximate value of the memory parameter $\tau_{0c}=14.70$, a chaotic crisis appears and the particle starts to transition between the two wells in an apparently random way. This is the value of the crisis-induced intermittency that we explore in what follows. For higher parameter values (beyond $\tau_0\approx16$) we find alternate region with appearance and disappearance of intermittency. 

In Fig.~\ref{fig:3} we have represented the invariant measures of the chaotic attractors of the first energy level (blue in Fig.~\ref{fig:1}) in the three main regions where intermittency can be observed (see again the bifurcation diagram in Fig.~\ref{fig:1}). These correspond to the values $\tau_0=8.1$, $\tau_0=15$ and $\tau_0=24$, respectively. In the former case, the particle spends large amounts of time exploring both wells, even though the average energy is below the local maximum of the double well. Correspondingly, in this case the intermittency phenomenon is not clearly visible. For the two latter values, we can neatly observe that the particle spends long transients inside each well with sudden and brief transitions between the two minima of the double well. The probability distribution in phase space is mostly concentrated near the minima of the wells (see Fig.~\ref{fig:3}). In particular, we study in detail the intermittency in the second example ($\tau_0=15$), which is the most robust of the three cases.

\section{Crisis-induced intermittency}\label{sec: Intermittency}

In order to study the intermittent dynamics, we first show time series, so that this phenomenon is easily illustrated. In Fig.~\ref{fig:46}(a) we can see a particle that experiences chaotic oscillations within the well at a value of the memory parameter less than the critical value of $\tau_{0c}=14.7$, at which the intermittency crisis takes place. Then, beyond the crisis, the time series is shown in Fig.~\ref{fig:46}(b) for $\tau_0=15$. We can identify how the system spends transients of variable length $\Delta t$ inside a given well, and then suddenly ``hops'' to the other well. The duration of the transients in each well vary unpredictably. To obtain the probability density of the residence time in each well between transitions, we have computed very long trajectories. Computer memory overflows have been avoided through the following technique. A long trajectory in some interval $[0, t_f]$ is computed. Then, since histories must be provided instead of initial points, the last part of the trajectory $[t_f-\tau_0, t_f]$ is saved and then interpolated using cubic splines to compute the next segment of the trajectory. Sometimes the particle inside one of the two wells hops over the barrier, briefly transitions the other well, and comes back to the initial well. Those transients that spend too small times (below $\Delta t = 100$) inside a well have been discarded, since they distort the statistics of residence times.
\begin{figure}
\centering
\includegraphics[width=0.9\columnwidth]{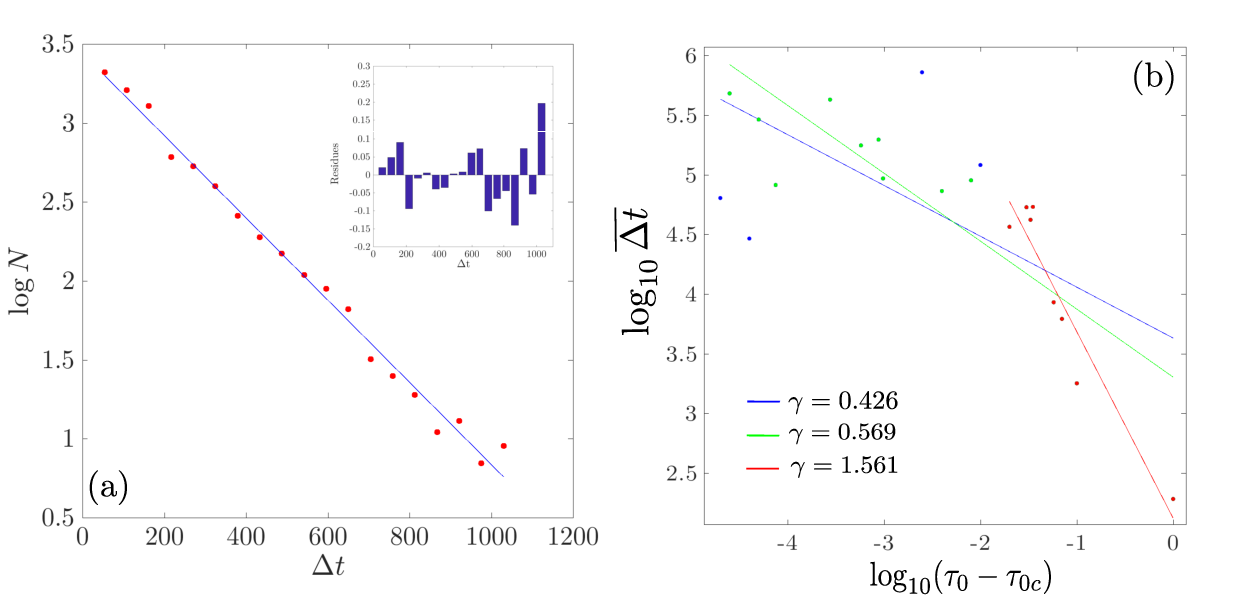}
\caption{\textbf{Residence times and scaling law}. (a) The distribution of residence time in each well for $\tau_0 = 15.0$, which follows an exponential distribution $P(\Delta t) \propto e^{-\Delta t/\delta}$, with $\delta=383.84$, characteristic of crisis-induced intermittency. (b) Scaling laws close to the crisis, where a power-law of type $\overline{\Delta t} \propto (\tau_0-\tau_{0c})^{-\gamma}$ is fitted, using a least squared fitting to different points. Several values for the critical exponent are estimated. The blue line is fitted using all the points, the green fitting uses the green and the red points, while the red least-squares fitting uses just the red points.}
\label{fig:6}
\end{figure}

As it can be seen in Fig.~\ref{fig:6}(a), the probability distribution of the residence times inside each well is nicely fitted by an exponential decay $P(\Delta t) \propto e^{-\Delta t/\delta}$, which is characteristic of crisis-induced intermittency~\cite{gre87, pom80}. The decay constant can be computed from the slope and has a value of $\delta = 383.84$. The residues are randomly distributed, as shown in the inset. This constant informs as about the average time spent by the particle inside the well, and it is a typical parameter computed in hyperbolic dynamical systems that scatter chaotically. It has also been investigated in works of hydrodynamic wave-particle entities~\cite{edd09}. In fact, recent works have shown a straightforward connection between systems of walking droplets and the paradigmatic chaotic Lorenz system~\cite{val22}. We highlight that the symmetry of the dynamical system, which had been broken by the retarded self-interaction, is now restored through crisis. This constitutes an example of \emph{symmetry restoring crisis}, which is one of the three fundamental mechanisms through which crisis-induced intermittency befalls in chaotic dynamical systems~\cite{gre87}. 
\begin{figure}
\centering
\includegraphics[width=1.0\columnwidth]{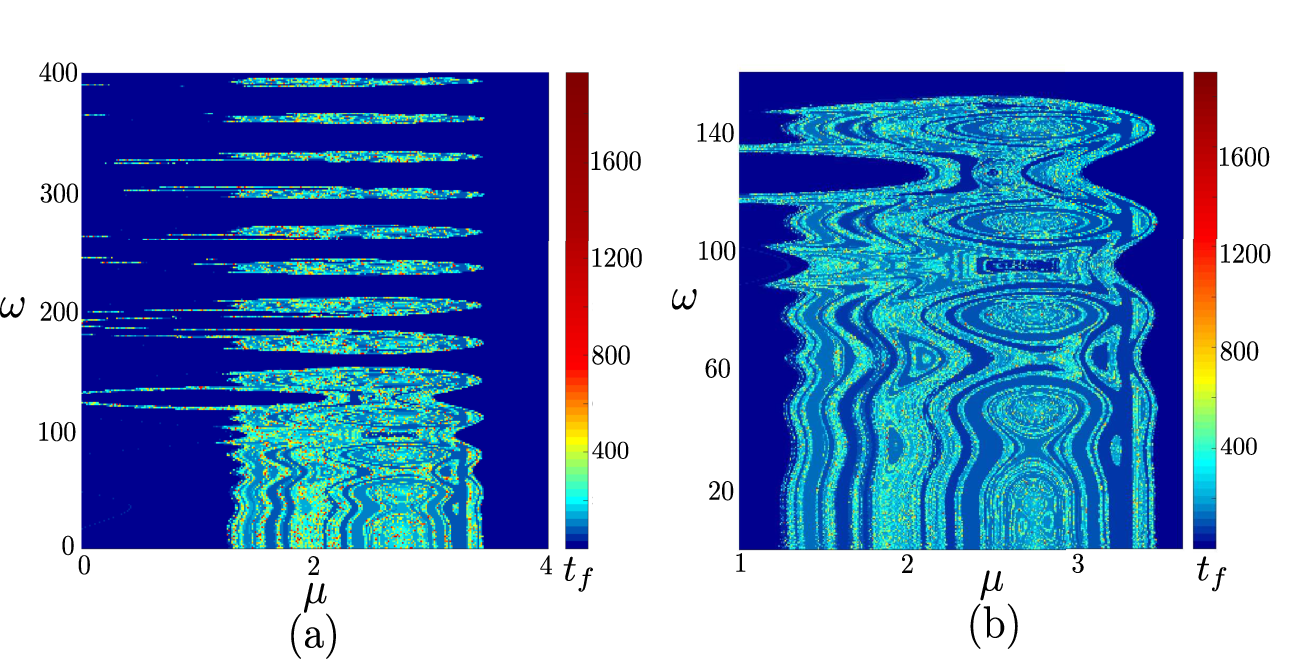}
\caption{\textbf{First passage times}. Colormap of first passage times $t_f$ for history functions $x(t) = \mu + A \sin(\omega t+\varphi)$ in the $(\mu,\omega)$ parameter space of initial conditions, with $\varphi$ approximately zero. The fractal nature of the distribution of the first passage times is clearly evident. (a) For a fixed value of the initial speed $\omega$, we observe great oscillations in the first passage times depending on the initial location of the particle inside the well fixed by $\mu$. A series of alternating islands are appreciated as we fix the initial position using $\mu$ and vary the initial speed using $\omega$. (b) A magnification of the plot in panel (a).}
\label{fig:7}
\end{figure}

\begin{figure}
\centering
\includegraphics[width=1.0\columnwidth]{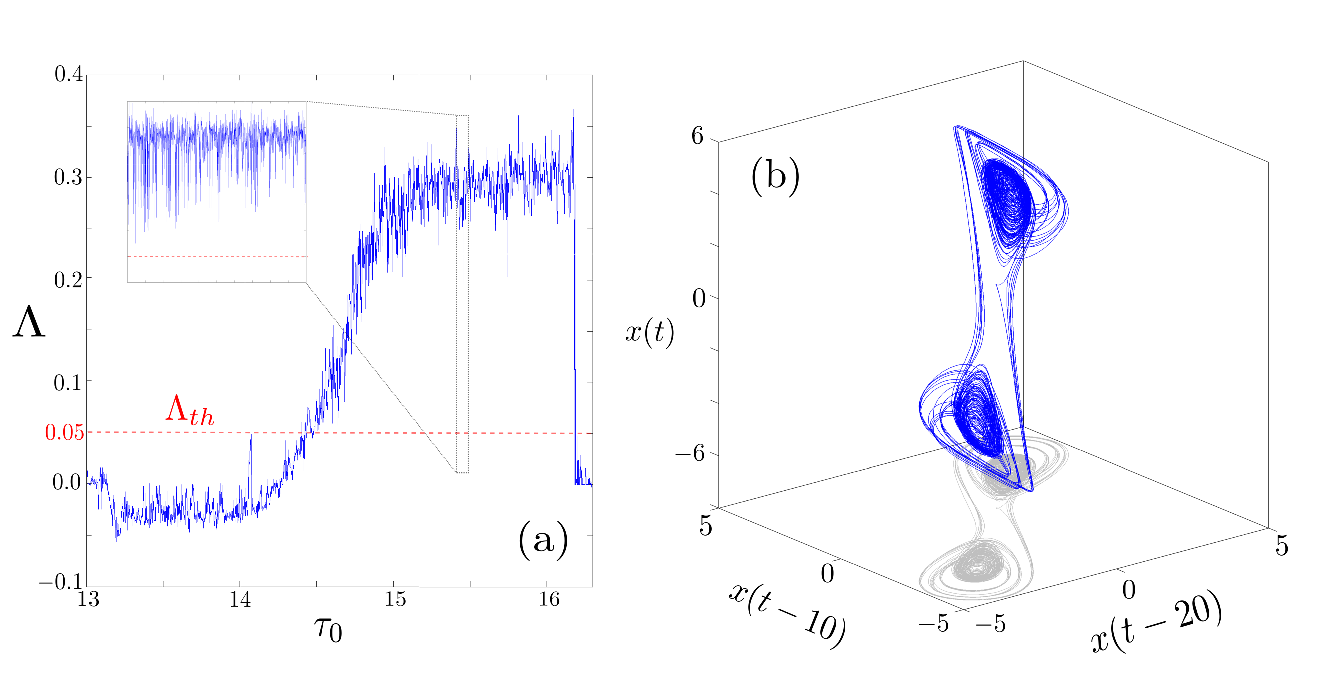}
\caption{\textbf{Largest Lyapunov exponent}. (a) The largest Lyapunov exponent $\Lambda$ is shown in the region where the crisis occurs as a function of the memory parameter $\tau_0$. A value of $\Lambda_{th}=0.05$ has been adopted as a threshold to determine whether the motion is chaotic (red dashed line). We see that the positive value of LLE mostly stays above the threshold, hinting at robustness of the intermittency to parameter perturbations in this range. A zoom in inside regions where the LLE exponent fluctuates are shown. (b) The 3D embedding for $\tau_0=15.5$ used to compute the Lyapunov exponents, showing a double-scroll attractor.}
\label{fig:8}
\end{figure}

Crisis induced intermittency is also characterized by studying the laws that govern the critical slowdown of the residence times as the parameter value gets close to the critical parameter value for crisis. For dissipative dynamical systems, the most frequent situation is a universal scaling behavior following a power law, which in our case would read $\overline{\Delta t} \propto (\tau_0-\tau_{0c})^{-\gamma}$. We have encountered computational difficulties in obtaining a single critical exponent, since the residual control integrator is inefficient, which prevents obtaining good statistics in reasonable time very close to the critical point. For this reason, we have estimated multiple critical exponents using points that are further away from the critical value of the maximum memory parameter $\tau_{0c}$. Specifically, we have considered points in the range $10^{-5} \leq \tau_0-\tau_{0c} \leq 1$, and obtained three estimates of critical exponents, which give us a range of possible values that are compatible with the theory of crisis-induced intermittency. As shown in Fig.~\ref{fig:6}(b), we obtain three estimates of the critical exponent. The first value uses all the points computed (blue line), and is $\gamma=0.426$, slightly below the value of $0.5$ typically appearing in two-dimensional maps~\cite{gre87}. Then, a second value has been obtained discarding points with large fluctuations (green line), providing a higher value of $\gamma=0.569$, which is slightly above the value of $0.5$, and which might suggest a quadratic tangency of the unstable manifold of some unstable periodic orbit inside any of the two wells~\cite{gre87}. It appears from numerical simulations that the system's phase-space dynamics may be embedded in three dimensions, suggesting a slow manifold of this dimension (see Fig.~\ref{fig:8}(b)). Nevertheless, we stress that this is just a conjecture, since our system is infinite dimensional. Finally, just to provide a conservative overestimate, we have used (see red fitting in Fig.~\ref{fig:6}(b)) only those points that have good statistics. These points lie in the range $[10^{-2},1]$ and give an estimate value $\gamma=1.561$, which would indicate that the tangency and the embedding are of higher dimension than three~\cite{gre87}. Despite all the difficulties obtaining a single critical exponent, we emphasize that the feature of critical slowdown is robust as one gets near the critical parameter value of crises.

\section{Tunneling uncertainty}\label{sec: Tunneling}

Now, we turn to investigate into the nature of the unpredictability of the chaotic intermittent transitions between the two wells of the double-well potential. To explore this, we adapt techniques typically used in the studies of \emph{transient chaos} to the present problem. Frequently, when studying chaotic scattering in dynamical systems with escape events, the escape times from a certain potential well are computed, and their fractal distribution is characterized by means of the uncertainty dimension~\cite{fer20}. We adapt the computation of the uncertainty exponent for our system, but by using first passage times instead of escape times, with initial conditions distributed inside one of the wells. The concept of first passage times is customary in the study of stochastic processes, as for instance the Kramer's problem~\cite{kra40}. We define first passage as the event when the particle crosses the value $x=0$ for the first time to escape to the other well.

In Fig.~\ref{fig:7} we show colormaps of the first passage times in the parameter space of initial history parameters $\mu$ and $\omega$. We have used histories in the form $x(t) = \mu + A \sin(\omega t+\varphi)$ for $t<0$, which allows us to control the initial position of the oscillator inside a well at $t=0$ by simply varying the parameter $\mu$. Consequently, the velocity before $t=0$ can be computed as $\dot{x}=A \omega \cos(\omega t+\varphi)$, which entails us to use the parameter $\omega$ as the initial speed, since we fix $A=0.01$. The initial phase $\varphi$ has been set to almost at zero value, thus it can be neglected. As can be seen in the figure, a fractal emerges, as it frequently appears in problems of chaotic scattering, where the largest times spent in some scattering region of the phase space correspond to points that are close to a chaotic saddle~\cite{fer20}. 
\begin{figure}
\centering
\includegraphics[width=1.0\columnwidth]{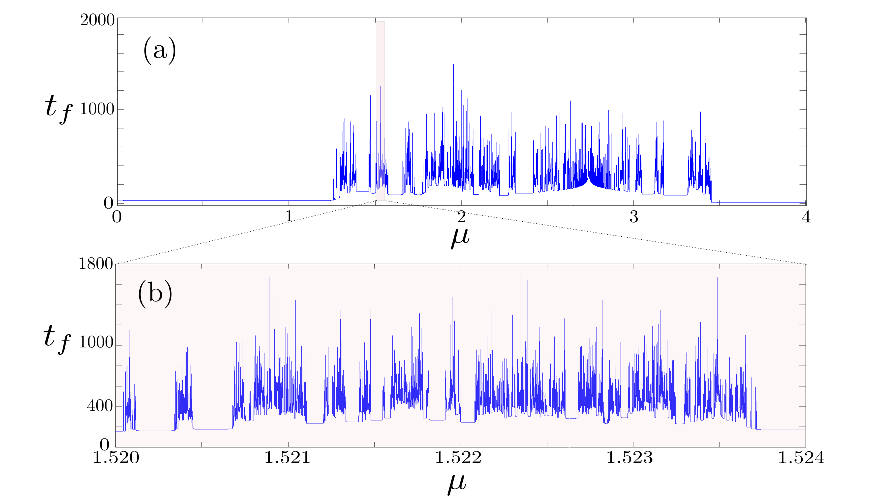}
\caption{\textbf{Cantor-like set}. The first passage times are plotted against the initial position $\mu$ of the particle inside the well, by setting the oscillator with very low speed ($\omega=0.01$) in the right well. (a) The chaotic nature of tunneling is manifest, and appears as a consequence of the fractal nature of this Cantorian set. (b) As we zoom inside the structure in the red window, the self-similarity of the first passage time is revealed.}
\label{fig:65}
\end{figure}

For a fixed value of $\omega$, a pattern of alternating regions inside the well with unpredictably changing first passage times is present. This is more clearly shown in Fig.~\ref{fig:65}(a), where Cantor-like set structure of the first passage times is manifest in the one-dimensional $\mu$ history space. This result might seen counter-intuitive at first glance, since one might expect that points closer to the boundary between the two wells to spend less time inside the well. However, this is not the case at all in chaotic dynamical systems, where the dynamics is mixed in the phase space similar to Smale's horseshoe. Interestingly, we see that as the initial speed $\omega$ of the particle is increased, a series of islands appear. This kind of behavior is reminiscent to the stability islands that frequently appear in the parameter space of time-delayed dynamical systems~\cite{lak11}. A blow up of the first island is shown in Fig.~\ref{fig:65}(b), where the fractal distribution first passage times is more clearly evident.
\begin{figure}
\centering
\includegraphics[width=0.55\columnwidth]{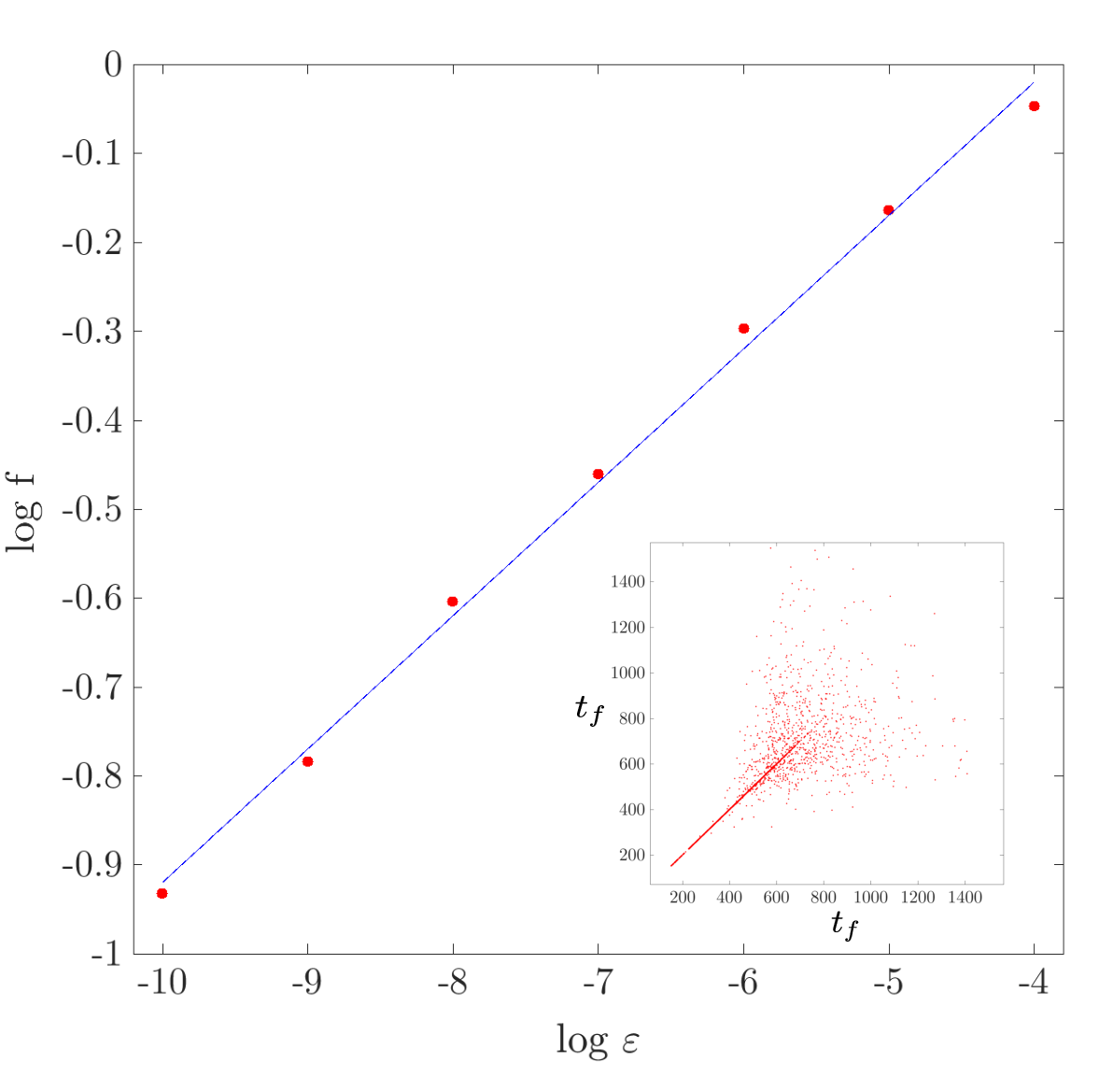}
\caption{\textbf{Uncertainty dimension}. A plot of the logarithm of fraction $f$ of events that are uncertain, against the distance between the points. Two points are considered uncertain if $|t_f(\mu)-t_f(\mu+\varepsilon)|<h$, with $h=1.0$. The uncertainty exponent $\nu$ can be computed from the relation $f(\varepsilon) \propto \varepsilon^\nu$, and gives a value of $\nu=0.1475$. The inset shows how the pairs of initial histories are related for a value of $\varepsilon=10^{-10}$. Many pairs gather in the diagonal, but still at such close distances a great number of points are far from the diagonal, revealing great uncertainty in the first passage times, which is due to chaos.}
\label{fig:5}
\end{figure}

We now derive the \emph{uncertainty exponent}, a number that allows us to estimate the fractal dimension of the Cantor set of the first passage times distribution~\cite{gre83,lau92}. In the present case, the uncertainty exponent measures the probability that two points with different initial histories in the phase space $\mu$ and $\mu +\varepsilon$, close to each other in one of the two wells, have a given difference in their value of their first passage times $|t_f(\mu)-t_f(\mu+\varepsilon)|$. Two points that exceed a certain small threshold $h$, are said to be uncertain. A scaling law can be then obtained by plotting the fraction of points that are uncertain for a range of distances in the initial histories of the randomly chosen pairs of points. The exponent of the power-law is called the uncertainty exponent, and measures how much the fist passage times fluctuate in the $\mu$ history space. Consequently, the fractality of the Cantor-like set can be characterized as one minus the value of the exponent, what gives the so-called uncertainty dimension~\cite{gre83}.

In the present computation we have used $9000$ pairs of points to compute the uncertainty exponent $\nu$ for each value of the scale $\varepsilon$, and simulating corresponding trajectories. Then, we plot the logarithm of the fraction of points that are uncertain, considering a value $h=1.0$, which is enough to ensure convergence. This value is three orders of magnitude smaller that the largest first passage times, and in general smaller than most values of the first passage time. As it can be seen in Fig.~\ref{fig:5}, we have obtained a nice least squares fitting of the power law, with randomly distributed residues. The resulting fractal dimension can be computed from the slope of this plot, and gives a value $D_{F} = 0.852$, which is above the dimension of the ideal Cantor set. This value shows that there is considerable uncertainty in the system, and that small variations in the setting of the initial conditions can produce great variations in the residence time inside one of the wells of our self-excited particle.

\section{Robustness of intermittency}\label{sec: Robustness}

We conclude our investigation by inspecting if the tunneling phenomenon persists across a continuous range of values of the memory parameter. Revisiting the bifurcation diagram in Fig.~\ref{fig:1}, we detect a range of values of the memory parameter between the critical values $\tau_{0c}=14.7$ and $\tau_0 \approx 16.15$, where periodic windows are not detected. We have zoomed in in this region in search for periodic windows, but we have not found them. This indicates that the tunneling phenomenon presents some degree of robustness, at least as compared with quadratic maps, where periodic windows densely appear in the chaotic domains of bifurcation diagrams. As far as we know, strictly robust chaos has been only demonstrated for non-smooth systems~\cite{ban98}. Whether strict robustness is present in high-dimensional continuous flows, such as the one at study, remains an open problem. Certainly, it would be desirable to have a quantitative measure that describes the density of periodic windows in the chaotic domain of a bifurcation diagram. But, from an experimental point of view, even if there are periodic windows, it might happen that noise blurs this windows below some experimental precision at a certain scale, what makes the concept applicable to a wide range of systems. 

As in Ref.~\cite{ban98}, in order to obtain some information about the degree of robustness, we compute the largest Lyapunov exponent (LLE). Because MATLAB’s inbuilt integrator does not allow to compute the LLE dynamically, we have taken advantage of embedology and used large time series. We follow a method introduced by Rosenstein et al. to efficiently compute the LLE from time series~\cite{ros93}. The computations have used an embedding dimension of $D=3$, and an embedding time-delay of $\tau = 10$. The mean period $T$ to compute the LLE considered can be obtained from spectral analysis (see Ref.~\cite{ros93}). We have used a value of $T=25$, which is an upper bound obtained for many parameter values of the attractor. The time of integration $t \in [0, 3000]$ has been considered and the maximum number of iterations for the algorithm was set to $1500$, adopting a conservative attitude (see again Ref.~\cite{ros93}). 

As shown in Fig.~\ref{fig:8}, the LLE grows quite rapidly as we enter the region of chaotic intermittency. The jump is not abrupt since chaos appears before the crisis phenomena. We have set a high threshold value $\Lambda_{th}=0.05$, above which the system is considered chaotic. Once the crisis takes place at $\tau_{0c}=14.7$ the LLE remains positive all the way up to $\tau_0=16.15$, where another crisis destroys the intermittent motion. The fluctuations in the values are considerable for this numerical algorithm. A blow-up around a peak where the exponents seems to fall down abruptly has been computed, showing that points do not cross below the $\Lambda_{th}=0.05$ threshold. In summary, we see that the intermittency phenomenon is present through a considerable range of ``continuous'' values, to the point that, from a practical point of view, we might expect that the phenomenon is not jeopardized by the existence of periodic windows within the chaotic regions. 

\section{Discussion and conclusion}\label{sec: Discussion}

We have shown that a damped particle in a double-well potential under the influence of time-delayed forces can show tunneling-like phenomenon between the two wells. This phenomenon would not be possible in this dissipative nonlinear oscillator without time-delayed history forces unless additional external forcing is present, e. g. external periodic forcing~\cite{gre87}. Similar phenomena have been studied in previous works with the Bogdanov-Takens resonance~\cite{coc18}, with constant linear time-delay. However, again non-homogeneous periodic drivings are required to produce transitions between the wells. Here, on the contrary, the system is \emph{self-driven} and the extent to which the memory effects are operational depends on the present dynamical state of the particle. This feedback through state-dependence is the key mechanism responsible for the complex dynamics. 

The observation of quantized energy states in our system is another interesting phenomenon. This phenomenon had already been shown in a simpler model with an external harmonic potential along with a harmonic retarded potential~\cite{lop23}. Similar to that study, here also we find that the maximum number of quantized states realized are two, as compared to having an infinite spectra of energy levels for quantum particles in confining potentials. This was believed to be a consequence of the neutral character of state-dependent delays in electrodynamics \cite{lop23}, which are present in the speed and the acceleration of the particle (e. g. the Li\'enard-Wiechert potentials). However, it is also due to the fact that our time retardation depends on the particle position, since the Gaussian function decays very fast, it possibly suppresses the existence of quantized states of large extent, because the memory effects are lost rapidly far from the center of the potential. In a future work, it will be shown that the state-dependence of the time delay on the particle's speed can produce many orbits conforming a Bohr-like model. In such systems with many quantized states, the resonant interaction between external time-dependent impulses and a large number of quantized states may drive transitions between different quantized states and we will explore this as well.

Finally, we would like to comment on the potential of differential equations with delays in providing insights into the intricate dynamics of other microscopic phenomena. A Helmholtz potential can be used to replace the Duffing's double-well used here, which physically could entail us to study other processes involving the tunneling effect, as for example radioactive decay processes, as described by R. Gurney and L. Condon \cite{gur28}. Or to give another example, we can study the entanglement of particles. The synchronization between nonlinear oscillators with time-delays has been extensively studied in the past literature~\cite{lak11}. This allows the possibility of two distant particle's to ``entangle'' by jiggling in synchrony, which can produce non-trivial correlations between them. In the present system, the coupling can be mediated by state-dependent delay functions such that when two particles interact, the time-delay function depends on the coordinates of both of them. Similar synchronization phenomena has been shown in pairs of interacting in-phase bouncing droplets~\cite{val18}, where the two particles exhibit inline  oscillations as well as promenading motion in synchrony. However, we recall that the pilot-wave of walking droplets decays exponentially fast in space and time, which makes faraway droplets hard to synchronize in the absence of non-local boundary effects~\cite{pap22}. Therefore, the present model opens forefront possibilities to investigate new dynamical phenomena and quantum analogs in nonlinear self-excited systems with memory.

\end{document}